# CT Reconstruction from Simultaneous Projections: A Step towards Capturing CT in One Go

Sajib Saha[1], Murat Tahtali, Andrew Lambert, Mark Pickering

School of Engineering and Information Technology
University of New South Wales, Canberra, Australia

**Abstract**

This paper focuses on minimizing the time requirement for CT capture through innovative simultaneous x-ray capture method. The state-of-the-art CT imaging methodology captures a sequence of projections during which the internal organ movements may lead to poor reconstruction due to motion artefacts. Traditional CT scanners' minimize such effect by taking more projections than necessary. In this work we focus on an innovative CT capture method that captures projections simultaneously, promising super fast scans along with possible radiation dose reductions. While the simultaneous CT capture model has already been proposed in our earlier work "Multi-axial CT Reconstruction from Few View Projections" (*in SPIE Optical Engineering+ Applications, pp. 85000A-85000A. International Society for Optics and Photonics, 2012*) and "A New Imaging Method for Real-time 3D X-ray Reconstruction" (*in SPIE Medical Imaging, pp. 86685G-86685G. International Society for Optics and Photonics, 2013*),  in this work we enhance the model through better initialization along with prior smoothing before successive iterations of the iterative algorithms. We also elaborate the model considering different X-ray source/detector configurations. Results show that it is possible reconstruct a cross-section slice by considering only four angular projections. With eight projections, the reconstruction is further improved. The main promising matter about this method is that, all these projections (i.e. four or eight) can in principle be captured simultaneously, implying CT capture in one go just like a chest X-ray.

## 1 Introduction

Computed Tomography (CT) is a medical imaging technique, employing tomography created by computer processing [1], whereas tomography is the process of imaging a cross-section, it comes from the Greek word tomos, which means a section or a slice or a cut [2]. The central idea of computed tomography is to produce 2-D and 3-D cross-sectional images of an object from flat X-ray images. CT has become one of the most important modalities in medical imaging; but unfortunately, the radiation exposure associated with CT is considered as a critical spin-off. With respect to patients' care, the least possible radiation dose is demanded.

Existing Computed Tomography (CT) systems are vulnerable to internal organ movements. This drawback is compensated by extra exposures and digital processing. CT being a radiation dose intensive modality, it is imperative to limit the patient's exposure to X-ray radiation, if only by removing the necessity to take extra exposures.

In order to minimize the number of projections, the proposed modality is based on simultaneous X-ray capture through a pinhole array akin to optical light field imaging. A fine array of pinholes is positioned in front of the detector allowing the determination of the direction of incoming X-rays along with their intensities. The simultaneous CT capture model has already been proposed in our earlier work [3, 4]. In this work, we elaborate the model considering different X-ray source/detector configurations. ART and CS-based ART are used to reconstruct the 2D slice as in [3, 4], however a better initialization is imposed for based on customized back-projection [5]. We also analyse the effect of applying smoothing filter after each ART iteration. It has been pointed out by Prun *et al.* in [6] that such smoothing might produce better results in ART. In this work, we also evaluate the effect of such smoothing on the reconstruction quality produced with and without applying sparsity prior [7] on the simultaneous CT capture modality.

---

[1] Corresponding author. Email: Sajib.Saha@student.adfa.edu.au

# 2 Background

## 2.1 Light field imaging

Light field imaging, the concept underlying the plenoptic camera, goes back to integral photography pioneered by Lippmann [8] and Ives [9]. The plenoptic camera is in principle not much different from an ordinary camera, the main difference being that a microlens array is placed just before the imaging sensor. This array of tiny lenses allows the recording of the direction of light rays in addition to their intensities. The technical details of this imaging method have been well documented by the Stanford team led by Professor Levoy [10, 11] who succeeded in building a plenoptic camera using a high-end commercial digital camera and a dense lenslet array.

In order to replicate the concept of lightfield imaging in X-rays, we should have the equivalent of a microlens array that works at X-ray wavelengths. As most materials are practically permeable to X-rays, they pass through almost without refraction. Only very dense materials such as lead come close to refracting them at impractical thicknesses and with extreme absorption, rendering the beam useless. At the European Synchrotron Radiation Facility website, a report citing recent progress in focusing X-ray Beams [12], that refractive optics made of low-Z materials such as Beryllium, Carbon, Aluminium, Silicon were developed at RWTH Aachen University. Figure 1 shows such lenses made of silicon with a focal distance in the range of a few millimetres at hard X-ray energies.

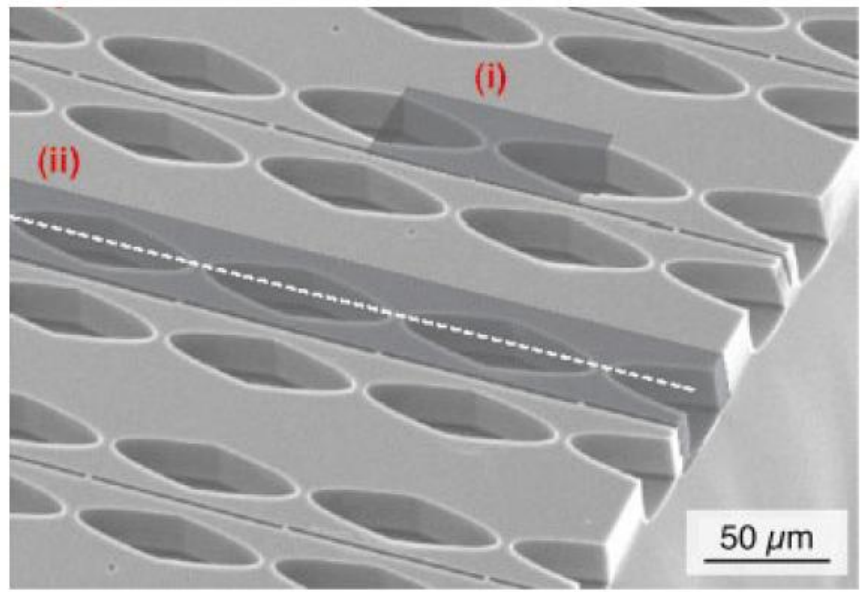

Figure 1: Scanning electron micrograph of an array of parabolic refractive X-ray lenses made of silicon. The shaded areas (i) and (ii) delimit an individual and a compound nanofocusing lens, respectively. (Reproduced from ESRF website [12]).

Another promising method is the more recent holographic or kinoform optical elements with the combination of refractive and diffractive properties. This method is claimed to eliminate drawbacks of purely diffractive or refractive elements and combine advantages like high transmission, absence of zero-order, high efficiency, etc... Figure 2 shows such focusing elements made of nickel by the Institute for Micro-Technology in Karlsruhe, using a lithographic process. This lens is reported to have a focal length of 4.5m at 212 keV X-ray energy.

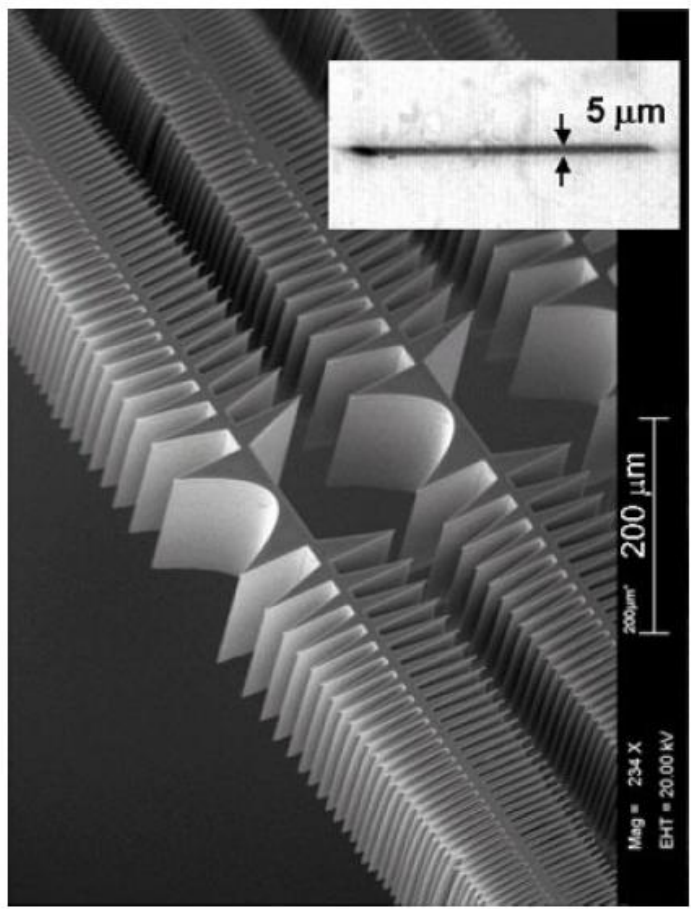

Figure 2: Scanning electron micrograph of a Ni kinoform lens with 140 single elements, each 300 μm high and 1500 μm wide. The lens is designed for 4.5 m focal distance at 212 keV X-ray energy. The insert (upper right) shows a 5 μm wide focus line measured using monochromatic 212 keV X-rays. (Reproduced from ESRF website [12]).

The above lenses were custom designed for synchrotron beamlines to produce concentrated light sheets of X-ray, therefore are linear lenses and their optical axes are parallel to the substrate. It is evident that neither of the above lenses can be used directly in our proposed plenoptic setup, however they are working proof that an X-ray lenslet array can be build using existing state-of-the-art technologies. It could be envisaged that a stacked array of these linear lenses can be built to form a 2D lenslet array.

A more readily available design is the classical Fresnel Zone Plate (FZP). NTT AT [13] from Japan is producing FZPs with minimum zone widths down to 25nm which would allow practical focal lengths of tens of centimetres in the hard X-ray energies.

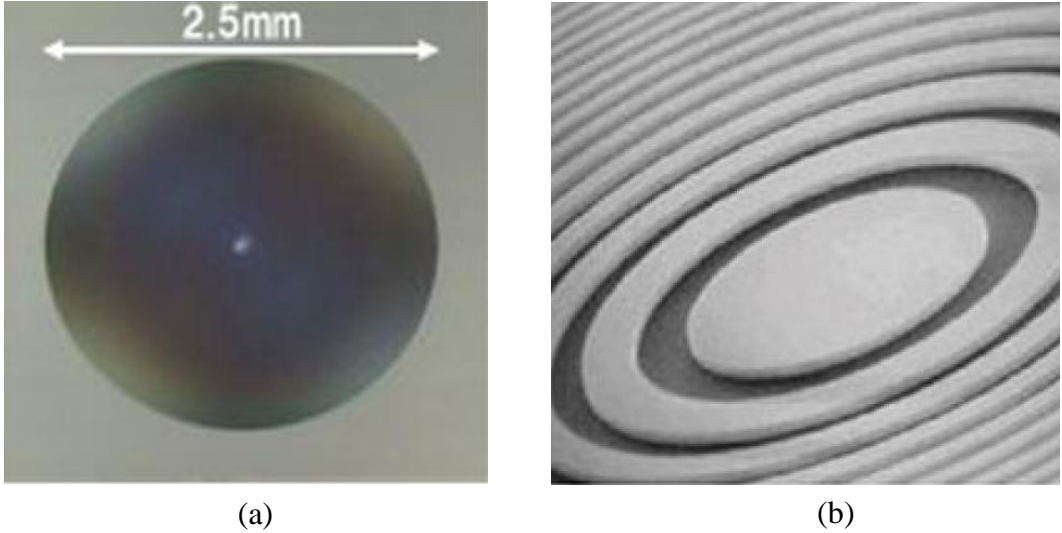

(a) (b)

Figure 3: (a) FZP for optical microscope and (b) SEM image of the same. (Reproduced from NTT AT website [13]).

Another company from the US, xradia [14] and the Center For X-Ray Optics at Berkeley Lab [15] are offering X-ray FZPs. Although these FZPs are offered as single lenses, there is no reason that the same lithographic etching techniques can't be used to produce an array of these lenses.

## 2.1 Algebraic Reconstruction Technique

Historically, the algebraic reconstruction technique was the first algorithm applied in CT [16], however filtered back-projection has been the method of choice by CT manufacturers [17] due to lower computational cost. Even though iterative approaches can be computationally expensive, efforts are being made to revisit iterative methods in [18-20] to ensure faster convergence of ART through the rearrangement of projections access order. Some interesting work has been proposed in [21-22] with regard to high computational capabilities available nowadays.

ART, which is considered as the core of iterative approaches, assumes that the cross section consists of an array of unknowns, and then sets up algebraic equations for the unknowns in terms of the measured projection data. In order to introduce the reader to ART, we will first show how we may construct a set of linear equations whose unknowns are elements of the object cross section. The Kaczmarz [23] method for solving these equations will then be presented.

In Figure 4, an imaginary square grid is superimposed on the image $f(x, y)$; where $f_j$ denotes the constant value in the $j$th cell. $N$ such cells represent the unknowns $f$ of the problem to be solved. A finite set of $M$ projections ($P=\{p_1, p_2,..., p_M\}$ ) is obtained, where each projection $p_i$ is defined by:

$$\sum_{j=1}^{N} w_{ij} f_j = p_i, \qquad i = 1, 2,..., M \tag{1}$$

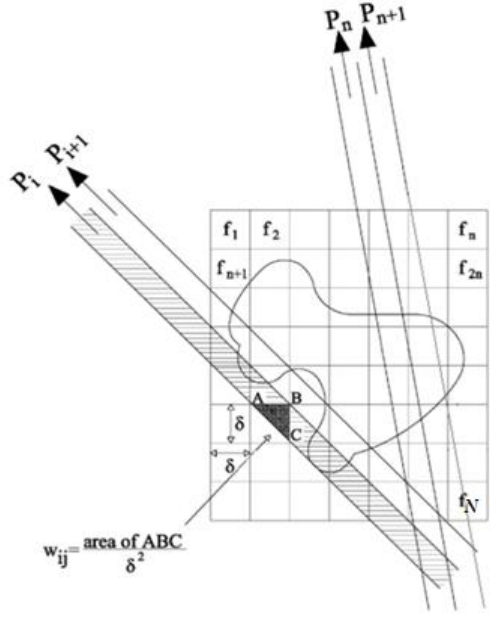

Figure 4: Image and projection representation for ART.

where, $w_{ij}$ is the weighting factor that represents the contribution of the $j$th cell to the $i$th ray integral as shown in Figure 4. $w$ is therefore a matrix of size $M \times N$. Thus, algebraic reconstruction algorithms, try to find a solution to the system of equations in (1). Conventional matrix theory methods can't be used to invert the system of equations in (1), since in reality $M$ and $N$ are significantly large and at the same time $M<N$ (under constrained). The Kaczmarz method [23] that finds out the solution is expressed by the following iterative equation:

$$f^{i,k} = f^{i,k-1} + \lambda \frac{(p_i - f^{i,k-1} . w_i)}{w_i . w_i} w_i \qquad (2)$$

where $i$ is the sub-iteration index from 1 to $M$, $k$ is the iteration index, and $w_i = (w_{i1}, w_{i2},...,w_{iN})$. Here, $\lambda$ is called the 'relaxation parameter' that controls the convergence rate of the algorithm.

## 2.3 Compressed Sensing

Compressed sensing (CS), also known as compressive sampling is a new technique being rapidly developed over the last few years. The theory states that when some prior information about the image (or signal) is available and appropriately incorporated into the image reconstruction procedure, an image can be accurately reconstructed even if the Shannon/Nyquest sampling requirement is not maintained [24]. In mathematical terms, if a vector $X$ contains at most $S$ nonzero elements and there are $K$ random measurements of $X$ such that $K \geq C.S.\log(D)$, where $C$ is a constant and $D$ is the dimension of $X$, then minimizing $L$-1 norm of $X$ reconstructs $X$ perfectly with an overwhelming probability.

Although the mathematical principal of CS is quite promising, its relevance in CT imaging relies on whether CT images are sparse or not. If an image is not sufficiently sparse, the CS algorithms will not be directly applicable to the problem. Fortunately, in CS theory, one can apply a sparsifying transform to increase the sparsity. The discrete gradient transform and wavelet transforms are frequently used for this purpose. Hence the basic idea of compressed sensing based image reconstruction can be summarized as follows: instead of directly reconstructing a target image, the sparsified version of the image is reconstructed. After the sparsified image is reconstructed, an 'inverse' sparsifiying transform is used to transform the sparsified image back to the target image. When the inverse-sparsifying transform is not explicitly available (such as for discrete gradient transform), iterative procedure is used to perform the inverse-sparsifying transform during the image reconstruction process [7, 25].

The compressed sensing idea has already been applied in CT [7, 25, 26, 27, 28, 29, 30], where successful reconstruction depends on proper initial guess, regularization parameter, algorithms used for sparse approximation, number of iterations and so on. In [25] it is shown that a local ROI can be exactly and stably reconstructed via the total variation (TV) minimization. The algorithm proposed in [7] minimizes the L1-norm of the gradient image as the constraint factor for the iteration procedure. In [26] an adaptive version of the original PICCS (the conventional CS objective function has been incorporated into the PICCS algorithm with a relative weighting) [30] algorithm has been proposed, that ensures higher image quality and reconstruction accuracy in regard to other approaches. It has been demonstrates in [27] that a small region of interest (ROI) within a large object can be accurately and stably reconstructed provided that a priori information on electron density is known for a small region inside the ROI. In [28], an approach for solving the CT interior problem based on the high-order TV (HOT) minimization, assuming that a ROI of piecewise polynomial was proposed. In [29], a numerical analysis framework, that the accurate interior reconstruction can be achieved on a ROI from truncated differential projection data via the TV or HOT minimization, assuming a piecewise constant (polynomial) distribution within the ROI, is shown.

## 3  Simultaneous CT Capture Model

Simultaneous CT capture model was proposed by us in [3, 4]. In this work, we consider 2 different setups of the model; setup I considers four angular projections at 0°, 45°, 90° and 135° and setup II considers eight angular projections at 0°, 22.5°, 45°, 67.5°, 90°, 112.5°, 135° and 157.5°. While the number of projections needed for setup II is twice the number of projections for setup I, the number of sources per projecton for setup I is 20 whereas for setup II it is 10.

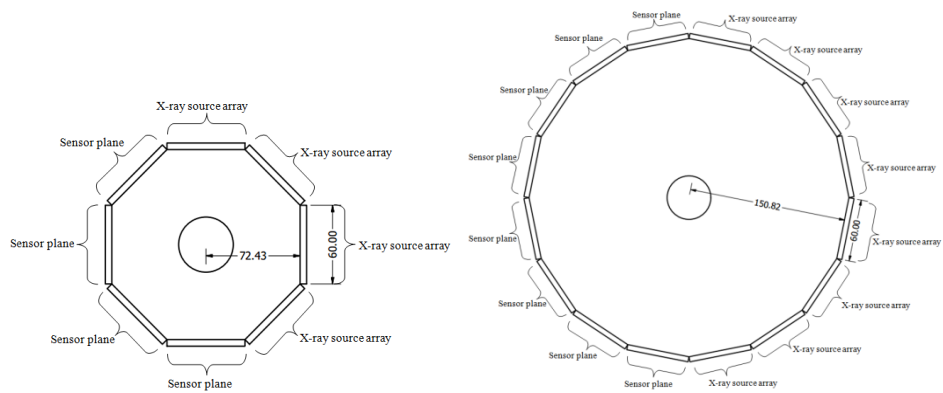

Figure 5: Schematic representation of the image capture modalities (a) considers four angular projections, (b) considers eight angular projections.

When working with more than one X-ray sources simultaneously, the directional information determining from which source any particular photon arrives at the detector needs to be obtained along with the photon intensity. An X-ray lenslet array could be used to this effect [3]. The only readily available means of producing the effect of an X-ray lenslet array was to use pinholes. For this, we used 0.4 mm tungsten plates with 50μm laser-drilled holes at 1mm pitch, which when placed over the X-ray camera, allowed the recording of 10 incident beams of different orientations in the x and y-direction, separately. It is worth mentioning that pinholes in conjunction with X-ray have already been used for astronomic observations by Stroke [31], where random pinholes where used to encode the aperture for later optical post-processing. Gamma cameras also use coded apertures [32].

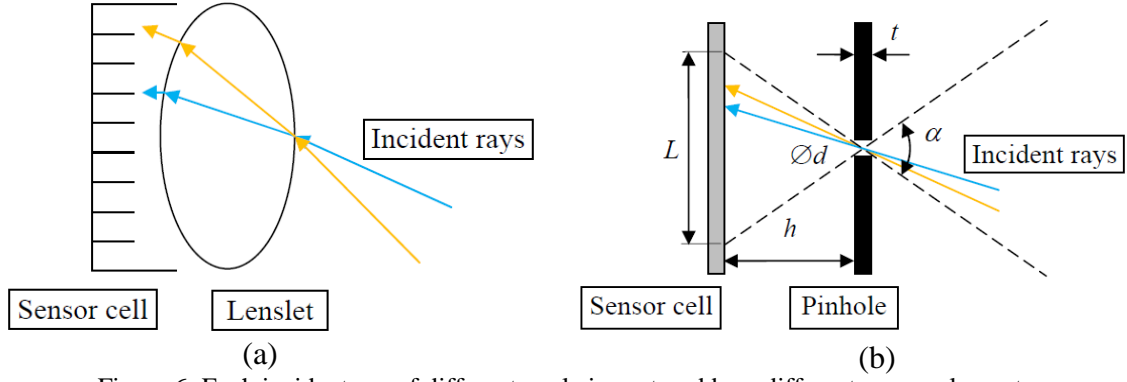

Figure 6: Each incident ray of different angle is captured by a different sensor element.
(a) refractive lenslet, (b) pinhole acting as a lenslet.

Figure 6 shows the refractive and pinhole versions of a lenslet, where *L* is the length of sensor area exposed under each pinhole as well as the pitch of the pinhole array, *d* is the pinhole diameter, *h* is the separation between the sensor plane and the pinhole array, *t* is the thickness of the masking material and *α* is the view angle, which is a function of the pinhole diameter and mask thickness. The view angle can be calculated from simple geometry as:

$$\alpha = 2\tan^{-1}\left(\frac{d}{t}\right) \qquad (3)$$

Each pinhole will form an image of diameter *L*, therefore the pitch of the pinhole array should be at least *L* in order to avoid overlap with the neighbouring cells, which can be calculated by:

$$L = 2\left(\frac{hd}{t}\right) \qquad (4)$$

Ideally, the distance *h* should be deduced by starting with the number of sensor pixels required under each pinhole, thus starting with *L*.

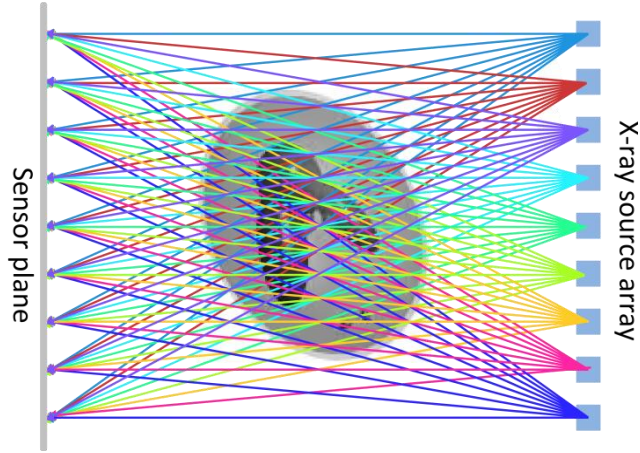

Figure 7: Simultaneous multiple fan-beam per projection.

Figure 7 shows a particular projection with simultaneous X-ray sources. Even though this imaging methodology is based on fan beam modality, it is slightly different in the sense that rather than covering the object completely, here the fan beam (generated from each of the X-ray sources) may cover the object partially, while at the same time there are overlaps between successive fan beams from different sources. The proposed imaging method assumes that each exposure produces simultaneously as many shadowgraphs as the number of X-ray sources. Figure 7 shows a close up of how such simultaneous shadowgraphs are encoded through the pinhole array. From Figure 7, without any fancy reconstruction but simple selection of corresponding pixels, it is possible to extract 2D X-ray representations of the object from the direction of the corresponding location of each X-ray source. One such exposure contains as many parallaxed images as the number of X-ray sources.

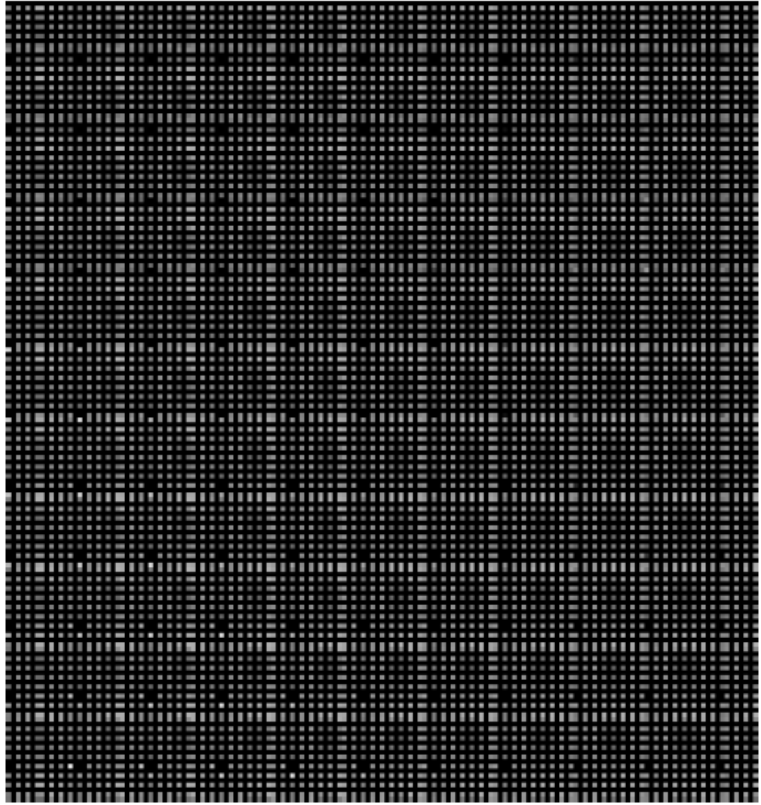

Figure 8: Close up of synthetically encoded multiple simultaneous projections exposure.

# 4 Reconstruction Method

Because of the non-uniform sampling of the projection data we are relying on iterative approaches for the successful reconstruction of the object. As the representative of iterative approaches, we are considering ART [23] and CS based ART [25]. Knowing each iteration of the iterative approaches comes at substantial computational cost, in order to achieve rapid convergence, we are motivated by Sajib *et. al.* [33] to start the iterative algorithms with a better initial guess. Likewise in [33], we also generate customized-back-projected image through dividing the summed value of the back-projected image by the number of rays that go through it (back-projected image). Once the customized-back-projected image is ready, we use it as the initial guess for ART and CS based ART. The pseudo-code for the considered reconstruction algorithms are as follows:

**Pseudo-Code: ART**

Step 1:
Compute customized back-projected image, $I_{Custz-back}$

Step 2:
$f^{ini} = T(I_{Custz-back})$
For $k = 1$ to $P_{ART}$ do
$\quad\quad f^0 = f^{ini}$
$\quad\quad$ For $i = 1$ to $M$ do

$$f^i = f^{i-1} + \lambda \frac{(p_i - f^{i-1}.w_i)}{w_i.w_i} w_i$$

    End For (*i*)

Step 3:

$$f_{m,n}^{temp} = T^{-1}(f^M)$$

    Smooth $f_{m,n}^{temp}$

$$f^M = T(f_{m,n}^{temp})$$

$$f^{ini} = f^M$$

End For (*k*)

$$f_{m,n}^{ART} = T^{-1}(f^M)$$

NB: Here $T(I_{2d})$ is a function that transforms $I_{2d}$ to a 1-dimensional vector, similarly $T^{-1}(I_{1d})$ transforms a 1-dimensional vector $I_{1d}$ of $m \times n$ elements to a 2-dimensional array of $m$ rows and $n$ columns.

## Pseudo-Code: CS based ART

Step 1:
Compute customized back-projected image, $I_{Custz-back}$

Step 2:
$f^{ini} = T(I_{Custz-back})$
For $k = 1$ to $P_{ART}$ do
    $f^0 = f^{ini}$
    For $i = 1$ to $M$ do

$$f^i = f^{i-1} + \lambda \frac{(p_i - f^{i-1}.w_i)}{w_i.w_i} w_i$$

    End For (*i*)

Step 3:

$$f_{m,n}^{temp} = T^{-1}(f^M)$$

    Smooth $f_{m,n}^{temp}$

$$f^M = T(f_{m,n}^{temp})$$

$$f^{ini} = f^M$$

Step 4:

$$f_{m,n} = T^{-1}(f^M)$$

    For $P_{CG} = 1$ to $P_{CG}$ do
        Compute the search direction $d_{m,n}$:

$$d_{m,n} = \frac{4f_{m,n} - f_{m+1,n} - f_{m-1,n} - f_{m,n+1} - f_{m,n-1}}{\mu_{m,n}} +$$

$$\frac{f_{m,n} - f_{m+1,n}}{\mu_{m+1,n}} + \frac{f_{m,n} - f_{m-1,n}}{\mu_{m-1,n}} + \frac{f_{m,n} - f_{m,n+1}}{\mu_{m,n+1}} + \frac{f_{m,n} - f_{m,n-1}}{\mu_{m,n-1}}$$

where $\mu_{m,n} = \sqrt{\frac{(f_{m+1,n} - f_{m,n})^2 + (f_{m,n} - f_{m-1,n})^2 + (f_{m,n+1} - f_{m,n})^2 + (f_{m,n} - f_{m,n-1})^2}{2\nabla^2}}$

$\beta = \max(|f_{m,n}|) \div \max(|d_{m,n}|)$

$f_{m,n} = f_{m,n} - \alpha \times \beta \times d_{m,n}$

$\alpha = \alpha \times \alpha_s$

    End For ($P_{CG}$)

$$f^{ini} = T(f_{m,n})$$

End For (*k*)

NB: Here $T(I_{2d})$ is a function that transforms $I_{2d}$ to a 1-dimensional vector, similarly $T^{-1}(I_{1d})$ transforms a 1-dimensional vector $I_{1d}$ of $m \times n$ elements to a 2-dimensional array of $m$ rows and $n$ columns.

Even though the iterative algorithms are computationally intensive, they are gaining popularity because of their interesting capability to produce meaningful result even with minimum number of and/or with non-uniform projections, supported nowadays with the wide spread availability of computational power [33].

## 5 Experiments and Results

The Shepp-Logan phantom was used for the simulation experiments. ART and CS based ART as described in section 4 were used to reconstruct the slice. Both for ART and CS based ART, the performance of smoothing was analyzed through applying Bilateral and Median filter separately. Figure 9 shows the intermediate reconstruction results of the Shepp-Logan phantom for the setup I, where we considered 4 angular projections at 0°, 45°, 90° and 135°.

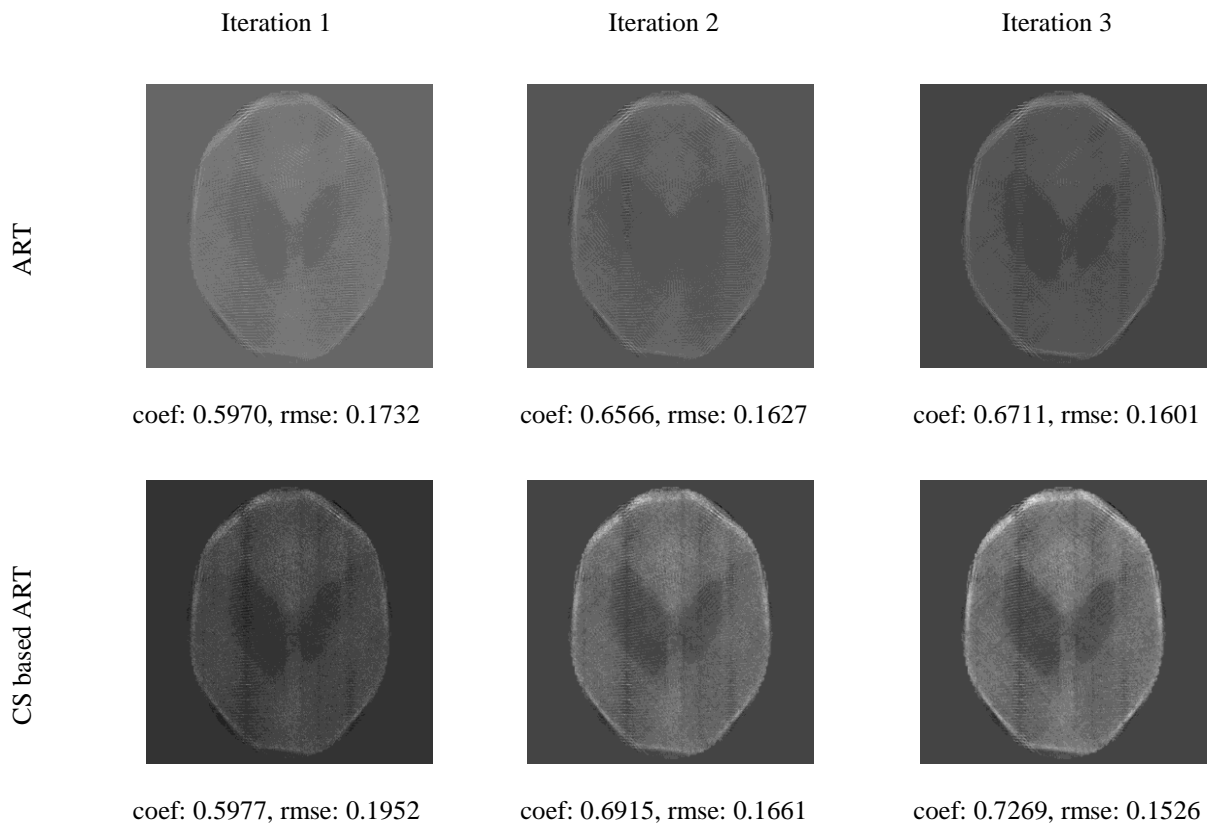

| | | | |
|---|---|---|---|
| ART (with bilateral filtering) | 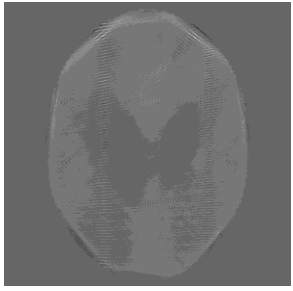 | 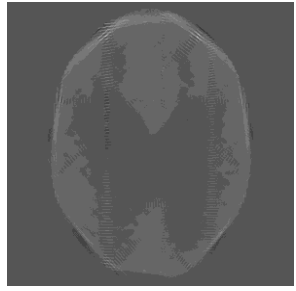 | 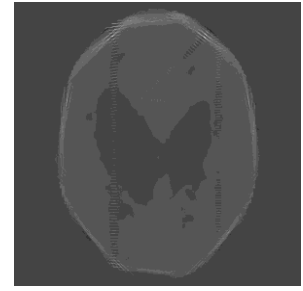 |
| | coef: 0.6416, rmse: 0.1645 | coef: 0.7076, rmse: 0.1513 | coef: 0.7250, rmse: 0.1474 |
| CS based ART (with bilateral filtering) | 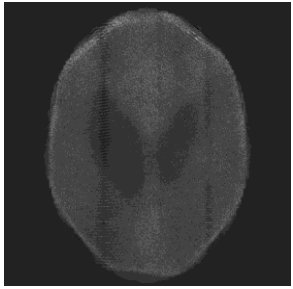 | 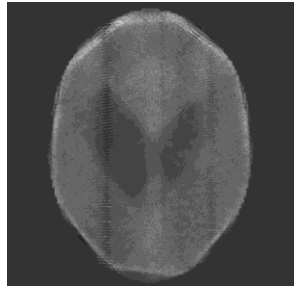 | 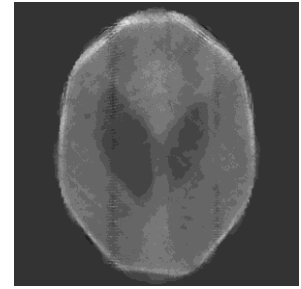 |
| | coef: 0.5963, rmse: 0.1936 | coef: 0.7102, rmse: 0.1623 | coef: 0.7484, rmse: 0.1486 |
| ART (with median filtering) | 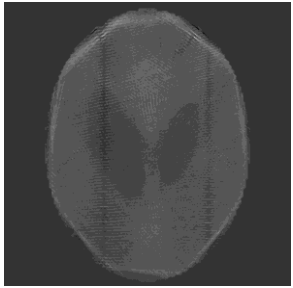 | 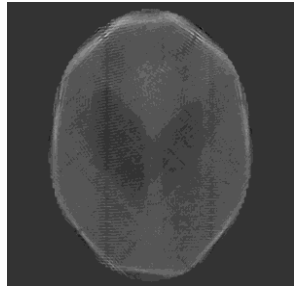 | 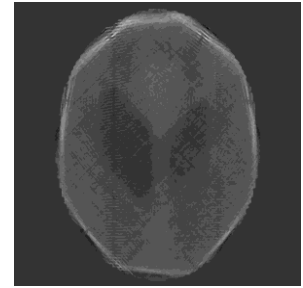 |
| | coef: 0.6875, rmse: 0.1578 | coef: 0.7521, rmse: 0.1420 | coef: 0.7718, rmse: 0.1366 |
| CS based ART (with median filtering) | 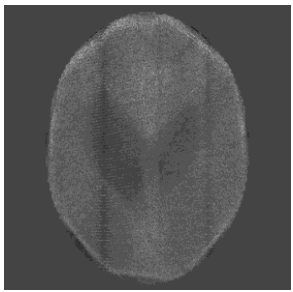 | 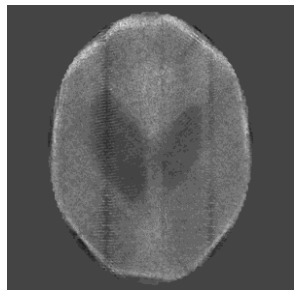 | 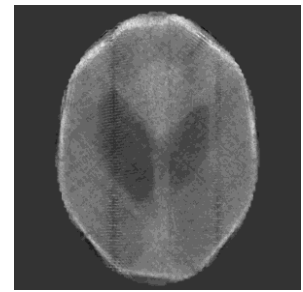 |
| | coef: 0.5827, rmse: 0.1950 | coef: 0.7050, rmse: 0.1631 | coef: 0.7450, rmse: 0.1489 |

Figure 9: Intermediate reconstruction of the Shepp-Logan phantom for setup I.

Figure 10 shows the intermediate reconstruction results of the Shepp-Logan phantom for setup II.

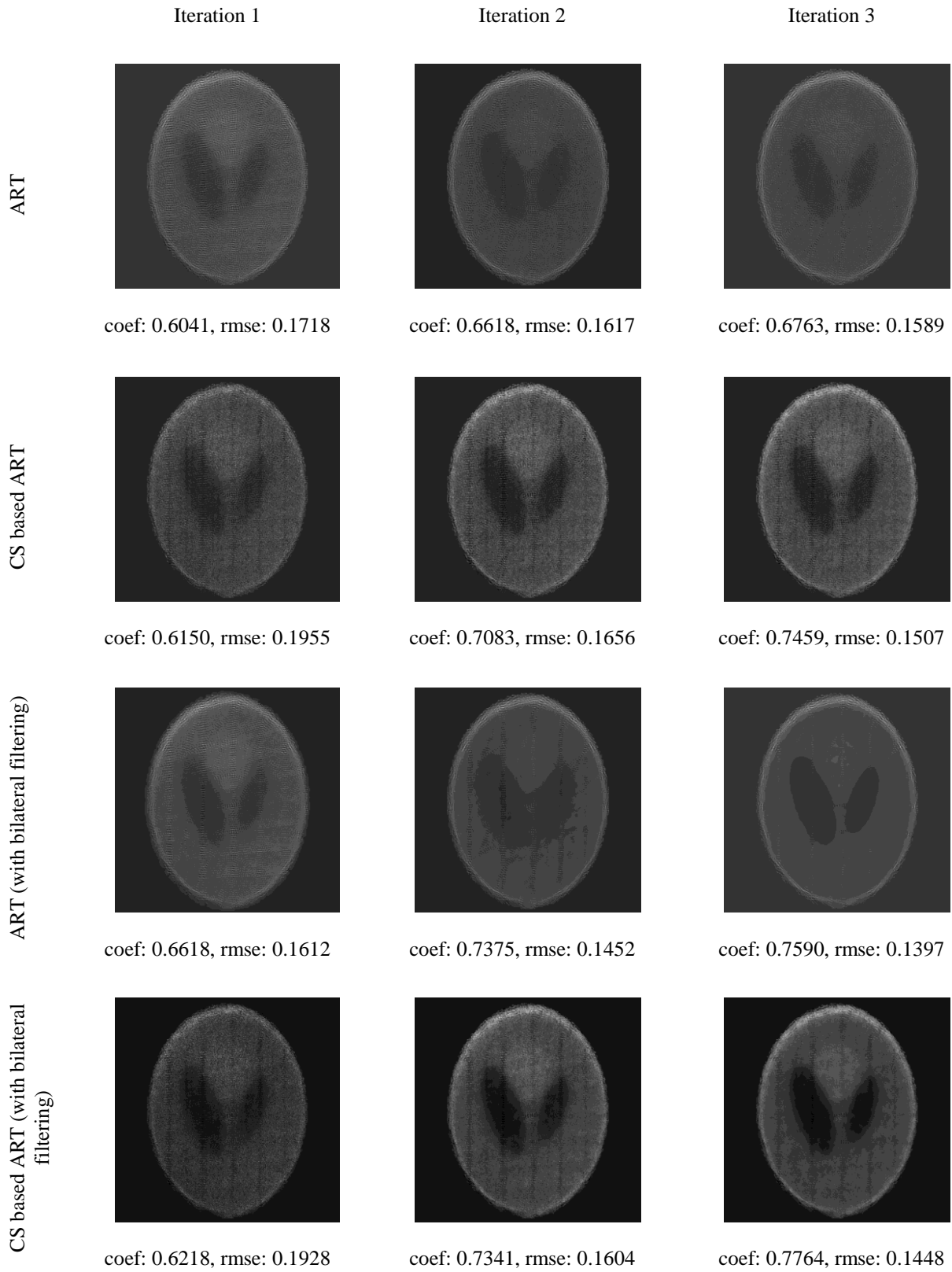

ART (with median filtering)

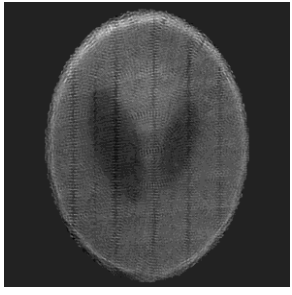 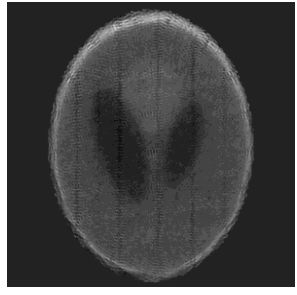 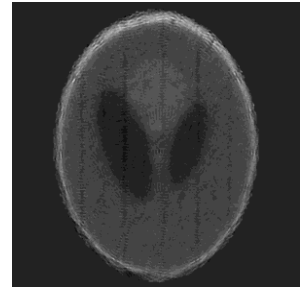

coef: 0.7100, rmse: 0.1548    coef: 0.7855, rmse: 0.1352    coef: 0.8151, rmse: 0.1261

CS based ART (with median filtering)

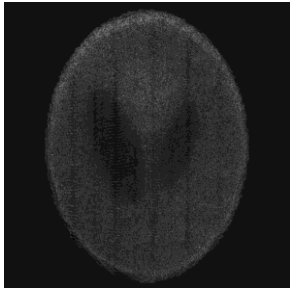 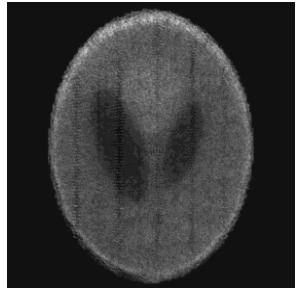 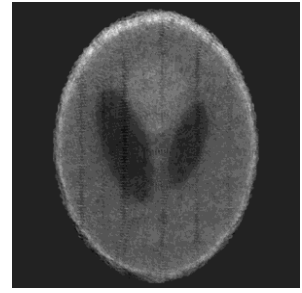

coef: 0.6046, rmse: 0.1941    coef: 0.7317, rmse: 0.1604    coef: 0.7753, rmse: 0.1443

Figure 10: Intermediate reconstruction of the Shepp-Logan phantom for setup II.

In order to evaluate the image quality achieved with successive iterations and with respect to different considered algorithms, we used correlation coefficients ($coef^{(k)}$) and root mean square error ($rmse^{(k)}$). The mathematical expressions of $coef^{(k)}$ and $rmse^{(k)}$ are given below.

$$coef^{(k)} = \frac{\sum_i (t_i - \bar{t})(x_i^{(k)} - \bar{x}^{(k)})}{[\sum_i (t_i - \bar{t})^2 (x_i^{(k)} - \bar{x}^{(k)})^2]^{1/2}} \qquad rmse^{(k)} = \left[\frac{\sum_i (x_i^{(k)} - t_i)^2}{\text{Number of Elements in } t_i}\right]^{1/2}$$

Where $t_i(\bar{t})$ and $x_i(\bar{x}^{(k)})$ represent the pixel (average) value in the original and *k*-th reconstructed images, respectively.

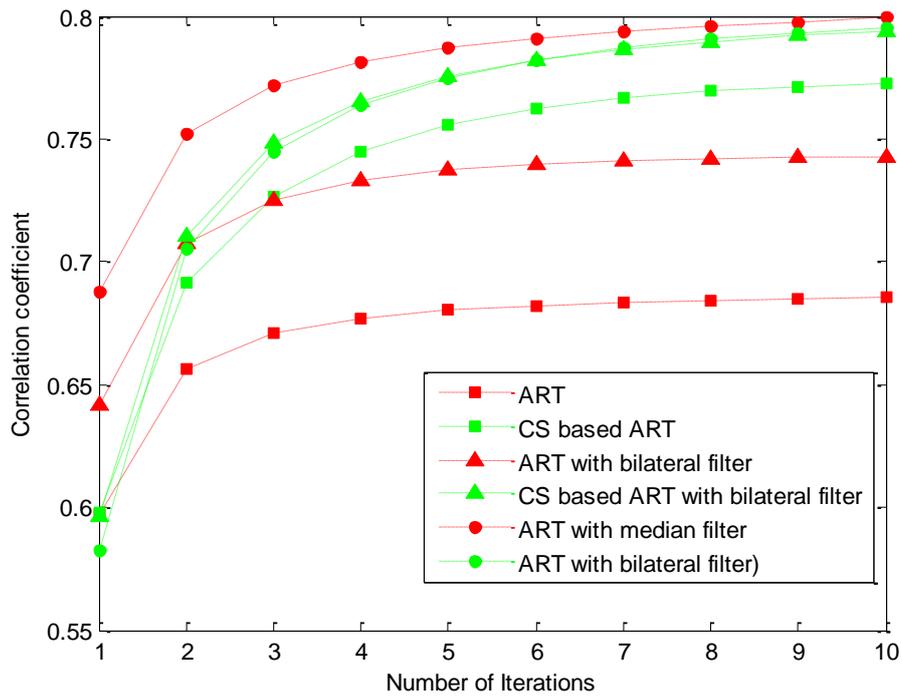

(a)

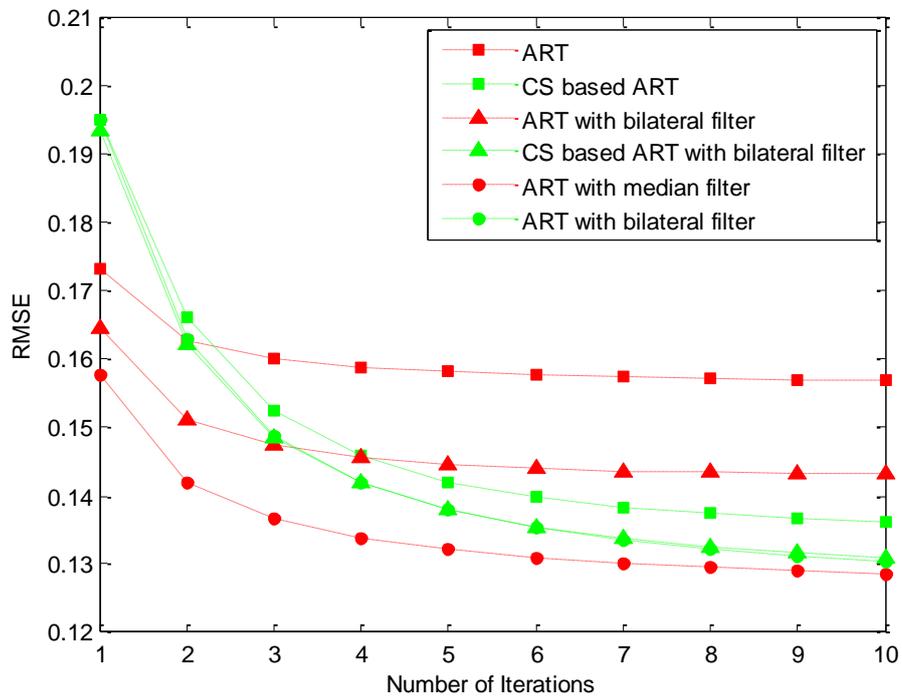

(b)

Figure 11: Mathematical evaluators of the reconstruction quality, (a) correlation coefficient (b) root mean square error (RMSE). Reconstructions were performed based on setup I.

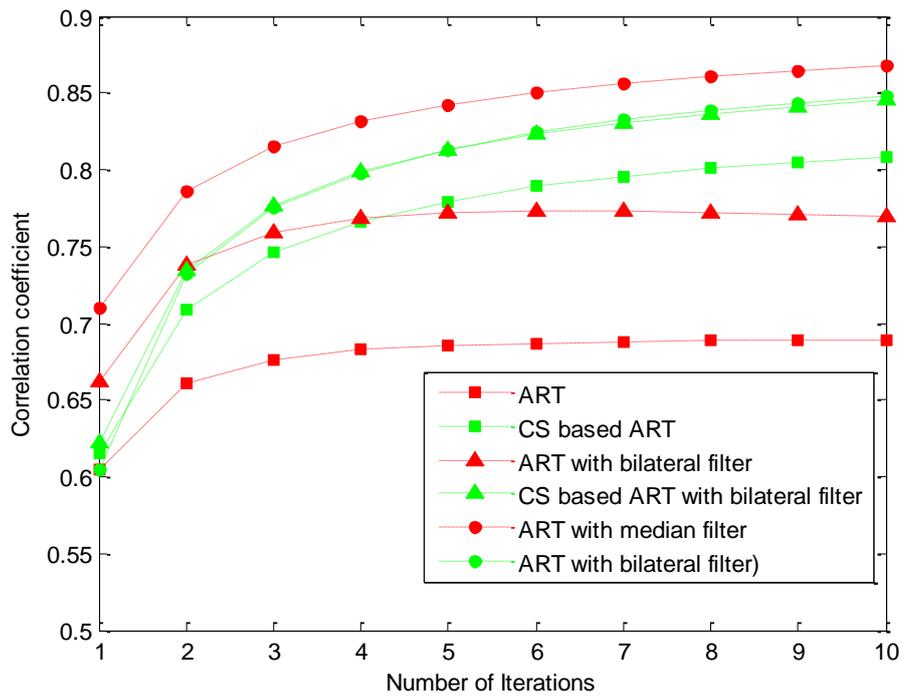

(a)

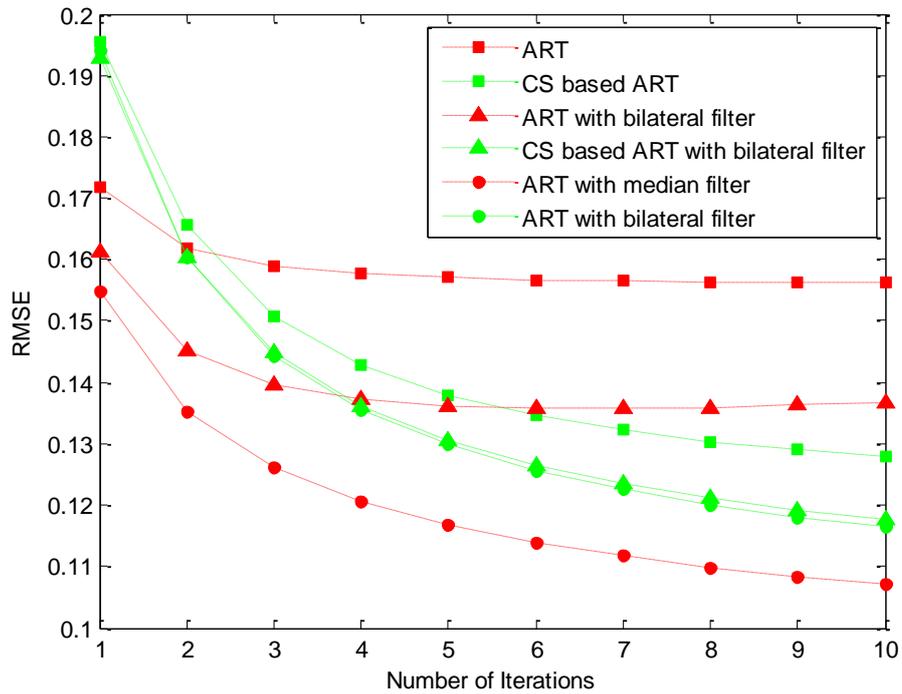

(b)

Figure 12: Mathematical evaluators of the reconstruction quality, (a) correlation coefficient (b) root mean square error (RMSE). Reconstructions were performed based on setup II.

From the results it can be observed that the application of smoothing filter at the end of each ART iteration ensures better reconstruction both for ART and CS based ART. Generally the performance of CS based ART is better than ART, however in the case of median filter, the performance of ART seem to be better than CS based ART, at least in the first few iterations. Setup II always outperforms setup I.

# 6  Conclusion and Discussion

We disclosed a novel CT image acquisition method along with reconstruction methodology to reconstruct the interior detail of an object. Customized back-projection along with smoothing filter has significantly contributed to the reconstruction quality produced by the simultaneous CT imaging modality by Sajib *et al.* in [3, 4]. Only 4 projections are required to reconstruct a slice of the cross-section, while traditional fan beam CT requires projections covering 360°. Four additional projections at 22.5°, 67.5°, 112.5°, 157.5° can produce visually more pleasant results, and supposedly more details. In principal all the projections (i.e. four projections for setup I or eight projections for setup II) can be captured simultaneously, thus the modality promises a drastic reduction of image capture time.

Even though the non-uniform sampling of the projections data forces the process to rely on computationally intensive iterative approaches for image reconstruction, the widespread availability of computational power nowadays makes this point moot. It also allows at the same time to exploit a wide range of advantages that iterative approaches offer over analytic ones. Even though the reconstructed image suffers from noticeable artefacts, we think that refinement and scaling up of this method is only an engineering challenge from this point on. The current implementation of ART has been limited to reconstructing slice data rather than direct 3D voxel reconstruction, which is computationally much more intensive, but potentially more efficient through the availability of more projection beams from a single exposure.

A technology developed at the UCLA, being further developed by Radius Diagnostics Research [34], promises the availability of affordable microemitter arrays that can be combined with our lightfield acquisition method. This has the potential of revolutionizing CT imaging, as truly real-time dynamic CT imaging of moving objects will become possible.

The imaging method described here is the subject of a patent application published under the Patent Cooperation Treaty (PCT) number WO2012155201.